\documentstyle[epsfig,aps,twocolumn,prb]{revtex}

\begin{document}
\bibliographystyle{prsty}
\draft

\title{Thickness dependence of the ground-state properties in thin films of the heavy-fermion compound CeCu$_6$}
\author{D.~Groten, G.~J.~C.~van~Baarle, J.~Aarts, G.~J.~Nieuwenhuys and
J.~A.~Mydosh} \address{Kamerlingh Onnes Laboratory, Leiden
University, 2300RA Leiden, The Netherlands}

\date{\today}
\maketitle
\section*{Abstract}
High-quality thin polycrystalline films of the heavy-fermion
compound CeCu$_{6}$ were prepared by sputter deposition. The
thicker of these films (with thickness up to around 200~nm)
reproduce the properties of the bulk compound CeCu$_{6}$. As the
thickness of the films is decreased, our measurements display
strong deviations from the bulk properties, namely, a suppression
of the heavy-fermion state. We show that possible ``external''
effects, like disorder, oxidation and morphology can be excluded
and that this size effect is therefore an intrinsic property of
CeCu$_{6}$. In addition, we investigate possible scenarios
explaining the size effect, and find that the proximity of
CeCu$_{6}$ to a quantum phase transition can account for this
striking result.

\section{Introduction}
Strongly correlated electron systems have been studied extensively
over the past two decades and produced a plethora of new and
intriguing results. Among these systems are exotic heavy-fermion
superconductors,\cite{steglich91,cox95} heavy-fermion paramagnets,
ferromagnets or antiferromagnets with tiny
moments,\cite{fisk88,nieuwenhuys95} spin glasses,\cite{mydosh97}
Kondo insulators\cite{aeppli92} and non-Fermi
liquids.\cite{maple95} Heavy-fermion compounds are often referred
to as the ``concentrated limit'' of dilute Kondo alloys, because
the local spin fluctuations arising from the hybridization of the
$f$ electrons with conduction-band states have essentially the
same origin as has the single impurity Kondo effect. The
Kondo-lattice concept can consistently explain many of the bulk
properties of Ce-based heavy-fermion compounds, in particular the
formation of a narrow coherent heavy-fermion band with $f$
symmetry at low temperatures.\cite{bauer91} It also provides a
single scaling temperature $T^\ast$ characterizing the properties
of the system at low temperatures. Now an intriguing question
arises: What length scales should be associated with (i) the
coherent quasiparticle state and (ii) the correlations arising
from the single impurity Kondo effect? In other words, will a
heavy-fermion system keep its bulk properties when one of more
dimensions are reduced?

In this paper, we address these queries by studying the transport
properties of thin films of CeCu$_6$ as a function of their
thickness. CeCu$_6$ is an archetypal heavy-fermion system with one
of the largest known electronic specific heat
coefficients\cite{stewart84a,schlager93} and no magnetic or
superconducting transition down to
20~mK.\cite{ott85,amato87,tsujii00} This makes it an ideal system
to study the ground-state properties of the heavy-fermion state.
Furthermore, it has attracted great attention in the past few
years due its proximity to an antiferromagnetic quantum critical
point (QCP). When doping with gold (CeCu$_{6-x}$Au$_x$) non-Fermi
liquid behavior is observed in CeCu$_{5.9}$Au$_{0.1}$ and an
antiferromagnetic ground state is found for
$x>0.1$.\cite{lohneysen96} Our main result is that as the
thickness of the films is decreased, the heavy Fermi-liquid state
is suppressed, and the behavior becomes that of a dilute
Kondo-impurity system. This result can best be explained with an
increase of the characteristic temperature scale $T^\ast$ as the
thickness of the films is decreased. We argue that the proximity
of CeCu$_6$ to the QCP can account for a correlation length of the
quantum fluctuations of ca.\ 10~nm, which leads to the increase of
$T^\ast$ as the thickness of the films approaches this length.

\section{Experimental}
The thin films of CeCu$_6$, with thicknesses ranging from around
10~nm to around 200~nm, were prepared by co-sputtering under
ultra-high vacuum conditions from two targets of pure Cu
(99.99\,\%) and Ce (Ames Laboratory, 99.995\,\%). They were
deposited onto Si(100) substrates with a Si$_3$N$_4$ buffer layer
at a temperature of 350$^\circ$C, and subsequently covered in-situ
at room temperature against oxidation with a Si layer. Thickness
and composition of the films were determined by electron-probe
microanalysis and Rutherford backscattering (RBS). The films are
found to be homogeneous both in-plane and as a function of depth,
and have sharp interfaces with substrate and protection layer.
X-ray diffraction (XRD) measurements demonstrate that the films
are polycrystalline (orthorhombic structure) with lattice
parameters within 0.1\,\% of those found for bulk
samples,\cite{groten99} and show a slightly preferred orientation
of the $b$-axis in growth direction. As the thickness of the films
is reduced, no variations of the lattice constants, nor of the
preferred orientation, are observed. High resolution electron
microscopy (HREM) has enabled us to prove that the grain size is
the same in films of thickness 189~nm and 53~nm, and of the order
of 10~nm, independent of film thickness. Also our RBS measurements
have enabled us to put an upper limit of 2~at.\% on oxygen
content. Resistance measurements were performed on structured
samples in a four-probe geometry (50--200~$\mu$m wide and 2--3~mm
between the voltage contacts) in applied magnetic fields up to 8~T
and at temperatures down to 20~mK. We have shown in an earlier
publication\cite{groten99} that for a film of thickness 189~nm,
the measurements of resistance versus temperature and those of the
magnetoresistance mimic the properties of bulk samples,
establishing the formation of a heavy Fermi-liquid state at low
temperatures in such thick films.

\begin{figure}
\begin{center}
\epsfig{file=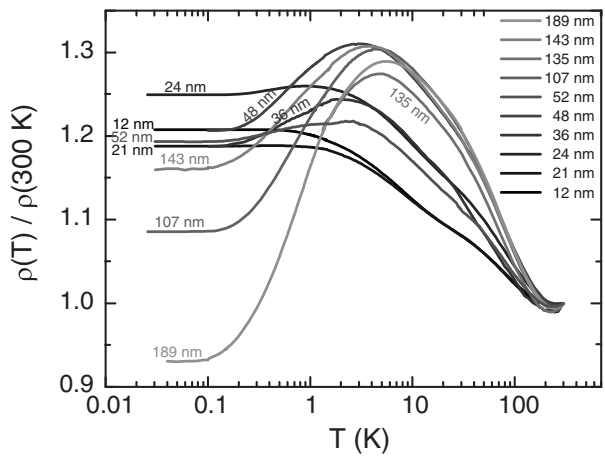, width=8cm}
\end{center}
\caption{Normalized resistivity $\rho/\rho(T=300~{\rm K})$ as a
function of temperature $T$ on a logarithmic scale in zero
magnetic field for CeCu$_6$ films of varying thickness. A darker
line means a thinner film.} \label{fig:resall}
\end{figure}

The resistivity data in zero magnetic field as a function of
temperature is shown for samples of various thickness in
Fig.~\ref{fig:resall}. The absolute values were normalized to
those measured at room temperature. This allows a better
comparison of the different samples, by removing uncertainties due
to the determination of the thickness ($\pm$10\,\%). Moreover, we
are not able to subtract non-magnetic contributions to the
resistivity, such as phonon contributions---this would require the
fabrication of LaCu$_6$ thin films. At high temperatures, the
logarithmic temperature scale enables us to compare these results
to a dilute Kondo system, in which the temperature dependence of
the resistivity is $-\ln T$. There is no clear range of
temperatures for the any sample in which this logarithmic behavior
is clearly observed. Crystalline electric field (CEF) effects
complicate the simple Kondo picture at temperatures of the order
of 50~K, due to the occupancy of higher lying CEF
levels.\cite{strong94} The general tendency at high temperatures
is that the slope $s=-d\rho/d(\ln T)$ is lower for thinner films
than for thicker films.

\begin{figure}
\begin{center}
\epsfig{file=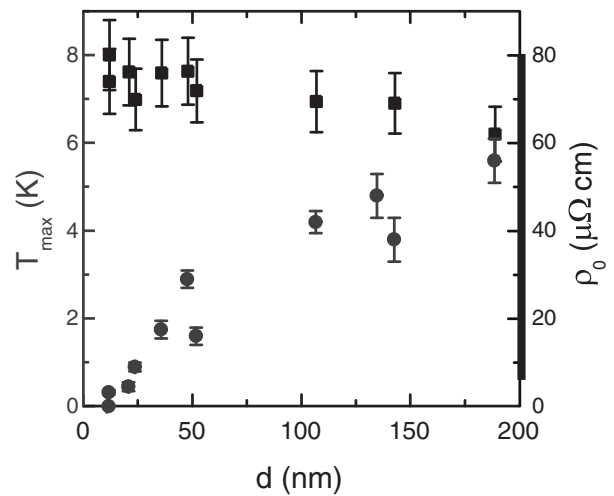, width=8cm}
\end{center}
\caption{Black squares, right axis: Residual resistivity $\rho_0$
of thin films of CeCu$_6$ measured at $\sim 30$~mK as a function
of film thickness $d$. The black bar on the right axis represents
the range of literature values found for bulk samples
(resistivities for the three crystallographic directions were
averaged for single crystals). Grey circles, left axis: Values of
$T_{\rm max}$ as a function of film thickness.} \label{fig:rho}
\end{figure}

At low temperatures, we see that large differences are found
between the thick films and the thin films. Whereas for the thick
films, there is a clear maximum of the resistivity $\rho$ at a
temperature $T_{\rm max}$ followed by a large drop upon lowering
the temperature, in contrast the resistivity just saturates for
the thinnest films. The absolute values of the residual
resistivity $\rho_0$ do not show any clear thickness dependence.
They are shown as a function of film thickness $d$ in
Fig.~\ref{fig:rho} (right axis). The black bar represents the
approximate range of values found in literature---averaged over
the three crystallographic directions for single crystals. Two
remarks can be made based on these measurements: (i) The
dependence of the residual resistivity on the thickness is very
weak, proving that no additional disorder is induced when the
samples are made thinner; (ii) The values of the residual
resistivity clearly lie in a metallic regime, well below the
Ioffe-Regel limit where the mean free path $\ell$ is equal to the
interatomic distance, which for CeCu$_6$ would give a maximum
resistivity of 300~$\mu\Omega$~cm for a metallic state. Also in
Fig.~\ref{fig:rho} are shown the values of $T_{\rm max}$ as a
function of thickness (grey symbols, left axis). The bulk values
found in the literature are as high as 15~K for particular
crystallographic directions. A clear tendency toward suppression
of $T_{\rm max}$ is found as the thickness of the films is
decreased.

\begin{figure}
\begin{center}
\epsfig{file=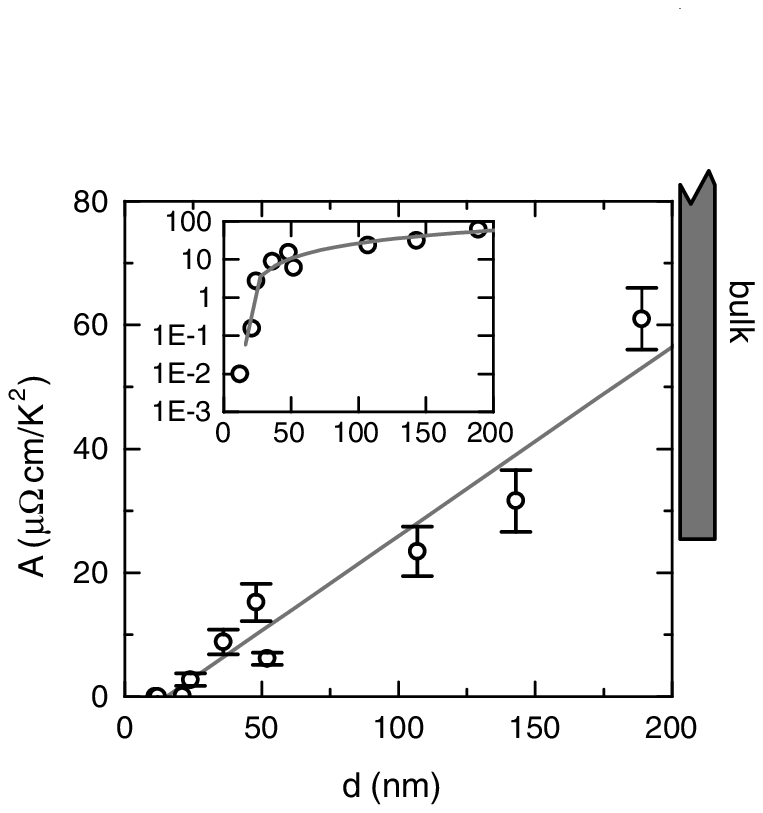, width=8cm}
\end{center}
\caption{Fermi-liquid coefficient $A$ of the low-temperature
resistivity as a function of sample thickness. The inset shows the
same points on a logarithmic scale. The grey bar signifies the
range of literature values for $A$ found for bulk samples. For
clarity reasons the bar is not extended to the maximum value of
112~$\mu\Omega$cm/$K^{2}$. For single crystals, values along the
three crystal directions have been averaged. The solid grey line
is a linear fit to the data. $A$ extrapolates to zero at $\approx
15$~nm.} \label{fig:acoeff}
\end{figure}

For $d>20$~nm, Fermi-liquid behavior is found\cite{groten99} at
very low temperatures ($T<200$~mK), $\rho=\rho_0 + AT^2$, where
$A$ is the Fermi-liquid coefficient related to the effective mass
$m^\ast$ of the quasiparticles by $A\propto
(m^\ast)^2$.\cite{baym91} The values obtained for $A$ are shown in
Fig.~\ref{fig:acoeff} as a function of film thickness. For
comparison, the values found in literature for bulk samples are
symbolized by the grey bar on the right side of the figure. There
is a dramatic decrease of $A$ as the thickness of the samples is
reduced. Below 50~nm this drop is better viewed on a logarithmic
scale, as seen in the inset.

Overall, the value of $A$ decreases by nearly four orders of
magnitude, which should be contrasted with a maximum difference by
a factor of five between the different bulk samples. For the
thinnest films of 11~nm and 12~nm, the resistivity just saturates
down to lowest temperatures, and no value of $A$ can be defined.
Our results clearly demonstrate that the {\em heavy} Fermi liquid
state is suppressed in our thinnest films.

\begin{figure}
\begin{center}
\epsfig{file=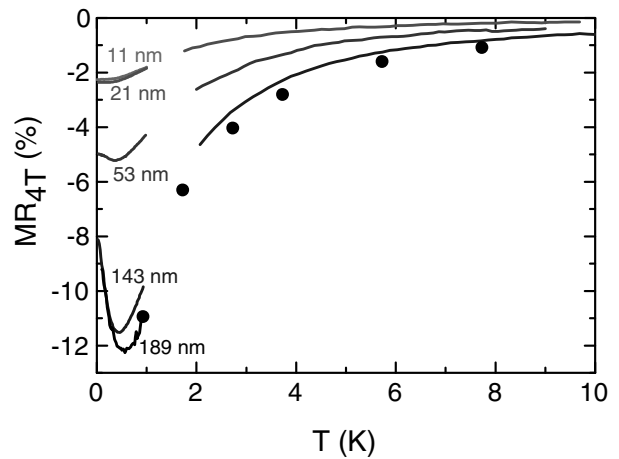, width=8cm}
\end{center}
\caption{Magnetoresistance MR measured in a field of 4~T as a
function of temperature for CeCu$_6$ thin films of varying
thickness. Only a few measurement points were taken for the film
of thickness 189~nm, represented by the black dots.}
\label{fig:mr}
\end{figure}

 The magnetoresistance (MR) is also strongly thickness
dependent: The magnitude of the negative MR decreases as the
thickness is decreased for all temperatures below 10~K. This is
shown in Fig.~\ref{fig:mr} where measurements of the MR in an
applied field of 4~T were performed. The MR is defined as ${\rm
MR} = (\rho(H)-\rho(0))/\rho(0)$. The thickest film (189~nm) shows
the typical MR of a Kondo lattice system, with an increasingly
negative MR as the temperature is lowered, followed by a minimum
and an increase of the MR which is due to the onset of coherence
and the HF state. In the thinnest films, the MR effect is much
weaker, and there is no minimum, proving again that the coherent
state is suppressed.

To summarize our experimental findings, we observe a dramatic
suppression of the heavy-fermion state in the ground state of
CeCu$_6$ as the thickness of the films is reduced. Whereas the
residual resistivity $\rho_0$ and the XRD measurements prove that
the intrinsic disorder is not changed as the films are made
thinner, characteristic signatures of the heavy-fermion ground
state such as the high Fermi-liquid coefficient $A$ and the sharp
minimum in the MR are reduced, and even completely suppressed in
our thinnest films. As mentioned previously, the quality of the
thin films was carefully analyzed, so that external effects such
as preferred orientation changes, the grain structure or oxygen
contamination can, as far as we can determine, be excluded as
possible explanations for the observed thickness dependence. This
striking effect is therefore an intrinsic property of CeCu$_6$,
and we will now consider possible scenarios for this
thickness-dependence property of the correlated state.

\section{Discussion}

As we have seen, the most common model used to describe
heavy-fermion systems is that of the Kondo lattice. Let us first
take this model as a starting point in our discussion and see
whether reducing the sample dimensions can lead to our results. We
can approach this in two ways: The {\em single impurity} Kondo
scattering is affected by a reduction of the size of the samples
(scenario I), or the Kondo temperature itself is size dependent
(scenario II), affecting the electron-electron interactions
leading to the coherent state in the Kondo lattice.

Finally, in scenario III, we will also put forward an alternative
way of looking at CeCu$_6$, based on the fact that this system
lies close to a quantum critical point.

Let us first remark that it is not simply the coherence that is
lost when the thickness is reduced. If that were the case,
CeCu$_{6}$ would just cross over from a periodic Kondo system to a
dense independent-impurity Kondo system. The resistivity would
increase down to zero temperature, with no change at high
temperatures, leading to a strong increase of the residual
resistivity. Also the magnitude of the MR would not change at the
minimum,\cite{kawakami86} as opposed to the large suppression we
observe for thinner films.

\subsection{Size dependence of single impurity Kondo scattering}

It is generally accepted by now that the original picture of a
``Kondo cloud'' with finite spatial extension of spin-polarized
electrons screening the magnetic moment of the Kondo
impurity\cite{gubernatis87} is not correct. This follows both from
theoretical considerations\cite{anderson70} and experimental
results.\cite{giordano96,chandrasekhar94,goldhaber98} However, a
number of experimental observations were made on the size
dependence of the Kondo effect, mostly using thin films or thin
wires of {\rm very dilute} Kondo alloys, e.g. {\em
Au}Fe\cite{giordano96,chandrasekhar94,chen91,blachly95,chandrasekhar94b}
or {\em Cu}Cr.\cite{haesendonck87,ditusa92} No clear picture
emerges from these experimental results, since in some cases a
size effect is indeed
observed\cite{giordano96,chen91,blachly95,chandrasekhar94b,ditusa92}
and in other cases not.\cite{chandrasekhar94,haesendonck87} What
these results have in common though, is that {\em the Kondo
temperature appears to be constant}. The size effect on the
transport properties, if any, shows a reduction of the Kondo
scattering term in thinner films or wires---{\bf scenario I}.

Two different theoretical models have been put forward to explain
these results. The first, by \'{U}js\'{a}ghy et
al.\cite{ujsaghy96} considers the effect of spin-orbit interaction
of the conduction electrons with the magnetic impurity. This
interaction induces an impurity-spin anisotropy close to the
surface of the sample, which reduces the effective thickness
contributing to the Kondo scattering of the conduction electrons.
While this may explain the observed reduction of the logarithmic
slope of the resistivity of CeCu$_6$ films as a function of
temperature at high temperatures, such a scenario cannot explain
our observation that at low temperatures, the Fermi-liquid
coefficient is completely suppressed in films of 10~nm thickness.
The second model, by Martin et al.\cite{martin97} describes how
the interplay between spin scattering (Kondo) and weak
localization (disorder) can modify the coefficient of the $\ln T$
dependence of the resistivity. However, their theory only applies
in the very dilute case, in which normal impurity scattering
leading to weak localization dominates over spin scattering.
CeCu$_6$ obviously does not belong to this category.

\subsection{Size dependence of the Kondo temperature}

Within the Kondo lattice model for heavy-fermions, the Kondo
temperature $T_{\rm K}$ is the fundamental energy scale of the
system. Other commonly used energy scales, such as the coherence
temperature or the RKKY temperature, can be directly related to
$T_{\rm K}$. Returning to the analogy with dilute Kondo systems,
let us consider a {\em thickness-dependent Kondo temperature} as a
possible cause of our results---{\bf scenario II}. From
Fermi-liquid theory, $A$ is proportional to $N(E_{\rm F})^2$, the
density of states (DOS) at the Fermi level squared. Our
measurements thus show that the DOS at the Fermi level is reduced
in thinner films. The Kondo temperature is related to the DOS at
the Fermi level via $T_{\rm K}\propto 1/N(E_{\rm F})$ provided the
Kondo interaction is much greater than the RKKY
interaction,\cite{doniach93} i.e.\ $JN(E_{\rm F})\gg 1$ with $J$
the magnetic coupling. An increase of the Kondo temperature
$T_{\rm K}$ with decreasing thickness is in qualitative agreement
with our results: An increased $T_{\rm K}$ reduces the magnetic
interactions at low temperature and therefore suppresses the heavy
Fermi-liquid state.

Unfortunately, transport measurements do not allow us to determine
unambiguously the Kondo temperature in our system. The maximum in
the resistivity $T_{\rm max}$ results from the onset of coherence
in the heavy Fermi-liquid state and is therefore not a good
measure for $T_{\rm K}$. The MR results can provide better
information about the Kondo temperature. The calculations by
Kawakami and Okiji\cite{kawakami86} show that for decreasing
$H/T_{\rm K}$, the amplitude of the MR decreases. The change of
amplitude in the MR of a factor of 5 as we observe in
Fig.~\ref{fig:mr} can be explained by an increase of $T_{\rm K}$
by about a factor of 4. As is the case for $T_{\rm max}$, we
cannot use the position of the minimum of the MR as a measure for
$T_{\rm K}$, since we cross over from a Kondo lattice to an
independent Kondo impurity system.

To explain the observed size dependence of Kondo scattering in
point-contact experiments also a variation of the Kondo
temperature with size was put forward.\cite{yanson95} The
theoretical explanation, given by Zar\'and and
Udvardi,\cite{zarand96} is not based on a decrease of the DOS but
rather on fluctuations of the local DOS near the surface of the
point contact. For very small point contacts, the contribution of
strong fluctuations of the local DOS near the surface of the point
contact leads to the observation of an enhanced ``effective''
Kondo temperature. The smaller the value of the bulk Kondo
temperature, the larger the enhancement in small point contacts.
We do not believe that such an explanation can hold for our thin
films of CeCu$_6$: Firstly, the effects we observe occur at very
large film thickness as compared to the length scales over which
one may expect fluctuations of the DOS.\cite{zarand96} Second, the
Kondo temperature of the compounds that were studied in the point
contact experiments were one to two orders of magnitude smaller
than that of CeCu$_6$, which has a bulk Kondo temperature of the
order of 1~K.

Whether scenario II can form the basis for explaining our results
is highly debatable. This scenario is based on a comparison with a
dilute Kondo model, with all its shortcomings when applying it to
a Kondo lattice system: One of them lies in the assumption of no
residual interactions or magnetic ordering appearing on the
periodic Ce lattice. Indeed, even for large $T_{\rm K}$,
interactions and antiferromagnetic order do remain a possibility
at low temperatures and thereby the transport properties would be
modified accordingly.\cite{tsujii00}

\subsection{Dimensional crossover due to quantum criticality}

\begin{figure}
\begin{center}
\epsfig{file=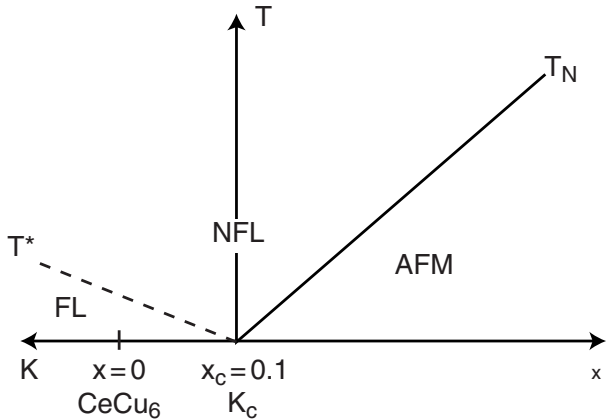, width=8cm}
\end{center}
\caption{Schematic phase diagram of the doped system
CeCu$_{6-x}$Au$_{x}$ which undergoes a QPT at $x = 0.1$. $K$ is
the coupling parameter measuring the strength of the quantum
fluctuations. (N)FL stands for (non) Fermi-liquid, AFM stands for
antiferromagnetic.} \label{fig:qcp}
\end{figure}

In order to get a deeper physical feeling for the observed
effects, let us take a step back, and attempt to describe
CeCu$_{6}$ in a more general manner, our {\bf scenario III}. We
already mentioned that CeCu$_{6}$ lies very close to a magnetic
instability. Whether or not it does actually order
antiferromagnetically at very low temperatures is still
debatable,\cite{schuberth95} but there is not much needed to
trigger such a transition: doping the system with gold leads to an
antiferromagnetic (AFM) ground state for CeCu$_{6-x}Au_{x}$ with
$x\geq 0.1$ (Fig.~\ref{fig:qcp}). It is now believed that the
non-Fermi liquid behavior observed at $x=0.1$ arises from the
quantum critical point (QCP) at that doping concentration. The
general picture is the following (see Fig.~\ref{fig:qcp}): When
describing the thermodynamics of a quantum system in $D$
dimensions, the expression for its partition function maps onto a
classical partition function for a system with $D+1$ dimensions;
the extra dimension is finite in extent and has units of
time:\cite{sondhi97} $\hbar /k_{\rm B}T$. At $T=0$, the quantum
system undergoes a quantum phase transition (QPT) when the
strength $K$ of the quantum fluctuations is changed and becomes
equal to a critical value $K_{\rm c}$: These fluctuations play the
role of the thermal fluctuations in the classical system.

Therefore, using our knowledge of critical phenomena, a spatial
correlation length $\xi$ diverges as $K\rightarrow K_{\rm c}$ with
a critical exponent $\nu$. Similarly, there must be a correlation
`length' $\xi_{\tau}$ in temporal direction, which may have a
different critical behavior than $\xi$ because of the possibly
different nature of the correlations in space and in time,
respectively. Quite generally, the relation between $\xi$ and
$\xi_{\tau}$ is governed by the dynamical exponent $z$,
$\xi_{\tau}\sim \xi^{z}$. The effect of finite temperatures is to
cut-off the low-frequency fluctuations via $k_{\rm B}T \sim
\hbar/\tau$. The characteristic temperature $T^{\ast}$ of the
system is the temperature at which the cut-off frequency $1/\tau$
is equal to the coherence frequency $1/\xi_{\tau}$. This
temperature marks a crossover with increasing temperature from a
$(D+1)$- to a $D$-dimensional system. Now, when the thickness of
the films is decreased, the spatial correlation length $\xi$ is
cut-off at the thickness $d$ of the film for sufficiently small
$d$. This leads to an effective increase of the coherence
frequency, since now $\xi_{\tau} \sim d^{z}$. For sufficiently
small $d$, the coherence `length' in temporal direction is now
directly related to the thickness of  the films, and therefore
also the characteristic temperature $T^{\ast}$ is related to $d$
for sufficiently small $d$:
\begin{equation}
\label{eq:t} T^{\ast} \propto d^{-z}.
\end{equation}

We have seen that it is difficult to determine the temperature
scale of our system based on transport measurements only. We
believe that the Fermi-liquid coefficient $A$ is the most physical
quantity that we can extract. Without requiring any of the
concepts of Kondo theory, we can still relate $A$ to $T^{\ast}$.
We know from Landau Fermi-liquid theory that $A\propto N(E_{\rm
F})^{2}$. Dimensional analysis requires that $T^{\ast}\propto
1/N(E_{\rm F})$. Our experiments suggest a linear relationship
between $A$ and the film thickness $d$. Putting these results in
Eq.~\ref{eq:t}, we obtain $z=0.5$.

To explain our results within the framework of quantum
criticallity, we sketched in Fig. 5 a phase diagram where $d^{-1}$
is put on the horizontal axis, which normally measures the
strength of the fluctuations. For, by decreasing the thickness of
the films, we are  increasing the characteristic temperature
$T^{\ast}$ thus moving away from the QCP. This suggests that
moving away from the Ne\'el ordered state enhances the quantum
spin fluctuations and therefore suppresses the heavy fermion
state, especially with a dimensionality reduction. In the phase
diagram, this is equivalent to moving towards the left on the
horizontal axis, or to stretching the $T$-axis and thereby pulling
up the line describing $T^{\ast}$. We do not observe a significant
increase of the residual resistivity due to this enhancement of
the quantum fluctuations.
 The relationship between $A$ and
$T^{\ast}$ was recently confirmed by measurements by Pfleiderer et
al.\cite{lohneysen99} They find that $A$ tends to diverge as
$T^{\ast}$ becomes zero at the QCP, proving that heavy-fermion
behavior is enhanced as one approaches the QCP. The value of the
dynamical exponent $z=0.5$ is quite unusual, since this would mean
that the spatial correlation length diverges faster than the
temporal correlation length as the QCP is approached. The expected
value of $z$ for an antiferromagnetic system is 2, and is
confirmed by the NFL behavior at the critical point, assuming
2-dimensional AFM fluctuations.\cite{rosch97,stockert98} We
therefore wish to stress here that the arguments we used to
determine $z$ were merely qualitative, and that in order to
confirm this picture of the HF system CeCu$_{6}$, measurements of
thermodynamic quantities such as the specific heat or the
susceptibility in these thin films would be more appropriate. They
are unfortunately very difficult to perform in view of the small
available volume in thin films.

A length scale $\xi$ of 10~nm might seem very large at first
sight. However, we are dealing with particularly ``slow''
quasiparticles, with a typical velocity $v_{\rm F}$ that is a
factor 1000 lower than free electrons in normal metals. Also we
have a system with a low characteristic energy scale $k_{\rm
B}T^\ast$ of about 1~K, which can be estimated, e.g., from the
temperature where the coherent state becomes important. This
characteristic energy corresponds to a time scale,
$\xi_\tau=10^{-11}$~s, leading indeed to a characteristic length
of $v_{\rm F} \xi_\tau=10$~nm. It is actually this very low value
of $T^\ast$ that is believed to make CeCu$_6$ and other so-called
``Kondo lattice'' compounds heavy-fermion systems.

\section{Conclusions}
Our measurements on thin films of CeCu$_6$ have exhibited a
dramatic change of the ground state properties upon approaching a
thickness of  the order of 10~nm.

Because CeCu$_6$ is usually described as a Kondo lattice system,
we have attempted to reconcile our results with the properties of
a {\em dilute} Kondo system. A size effect of the Kondo effect
itself ({\bf scenario I}), as observed experimentally in dilute
Kondo systems, does not account for the scaling of characteristic
energies that we observe as a function of thickness ($A$ and
$T_{\rm max}$). The characteristic energy of the Kondo lattice
model is the Kondo temperature $T_{\rm K}$, and a thickness
dependence of  $T_{\rm K}$ can qualitatively explain our results
({\bf scenario II}). However, it is not clear whether this
scenario also applies to a Kondo lattice, and why fluctuations of
the DOS should occur at the relatively large length scale of
around 10~nm.

The proximity of CeCu$_6$ to a quantum critical point has led us
to seek an explanation in terms of another description of CeCu$_6$
({\bf scenario III}) . The physics of a system near a QCP is
governed by two correlation lengths, $\xi_\tau$ in the time
direction, and $\xi$ in spatial direction, that diverge as one
approaches the QCP. When, as is the case for CeCu$_6$, the quantum
fluctuations are so large that the system is not ordered at $T=0$,
$\xi_\tau$ sets a characteristic temperature scale $T^\ast$ above
which the system ``loses'' one dimension. Because $\xi_\tau$ and
$\xi$ are coupled through the dynamical critical exponent $z$, an
effective reduction of one spatial dimension below $\xi$ has the
effect of increasing the characteristic temperature scale
$T^\ast$. This leads on the one hand to a ``stretching'' of the
temperature axis, i.e.\ a flattening of resistivity as a function
of temperature. On the other hand, it leads to a lower DOS at the
Fermi level, and hence to a weakening of the correlations
(decrease of $A$). This scenario therefore provides a  reasonable
``starting point'' for the length scale that was found in our
experiments although we can not explain the very small values of
$A$ for the thinnest films and the residual resistivity does not
reflect the increase of the quantum fluctuations.

\section*{Acknowledgements}
We wish to thank L.~Qin and H.~W.~Zandbergen from the National
Centre for HREM in Delft, The Netherlands, for the HREM
measurements on our films. Furthermore, we would like to
acknowledge the enlighting discussions J.~Zaanen on the theory of
quantum phase transitions. This work was supported by the Dutch
Foundation for Fundamental Research on Matter (FOM).

\end{document}